%harvmac
%\input harvmac

%%%%%%%%%%%%%%%%%%  tex macros for preprints, cm version %%%%%%%%%%%%%%
%                     (P. Ginsparg, last updated 9/91)
%                if confused, type `b' in response to query 
%
%---------------------------------------------------------------------%
%% site dependent options: 
%% \unredoffs and \redoffs define horizontal and vertical offsets 
%% respectively for unreduced and reduced modes. \speclscape defines
%% the \special{} call that sets printer to landscape (sideways) mode.
%% from standard set below, leave uncommented as appropriate or redefine
%
%%% next 400dpi
%\def\unredoffs{} \def\redoffs{\voffset=-.31truein\hoffset=-.48truein}
%\def\speclscape{\special{landscape}}
%
%%% apple lw
\def\unredoffs{} \def\redoffs{\voffset=-.31truein\hoffset=-.59truein}
\def\speclscape{\special{ps: landscape}}
%
%%% qms lasergrafix:
%\def\unredoffs{} \def\redoffs{\voffset=-.4truein\hoffset=.125truein}
%\def\speclscape{\special{qms: landscape}}
%
%%% saclay A4 paper:
%\def\unredoffs{\hoffset-.14truein\voffset-.2truein} 
%\def\redoffs{\voffset=-.45truein\hoffset=-.21truein} 
%\def\speclscape{\special{landscape}}
%
%---------------------------------------------------------------------%
%
\newbox\leftpage \newdimen\fullhsize \newdimen\hstitle \newdimen\hsbody
\tolerance=1000\hfuzz=2pt
\catcode`\@=11 % This allows us to modify PLAIN macros.
\def\bigans{b }
\def\answ{b }

%\message{ big or little (b/l)? }\read-1 to\answ
%
\ifx\answ\bigans\message{(This will come out unreduced.}
\magnification=1200\unredoffs\baselineskip=16pt plus 2pt minus 1pt
\hsbody=\hsize \hstitle=\hsize %take default values for unreduced format
\else\message{(This will be reduced.} \let\l@r=L
\magnification=1000\baselineskip=16pt plus 2pt minus 1pt \vsize=7truein
\redoffs \hstitle=8truein\hsbody=4.75truein\fullhsize=10truein\hsize=\hsbody
\output={\ifnum\pageno=0 %%% This is the HUTP version
  \shipout\vbox{\speclscape{\hsize\fullhsize\makeheadline}
    \hbox to \fullhsize{\hfill\pagebody\hfill}}\advancepageno
  \else
  \almostshipout{\leftline{\vbox{\pagebody\makefootline}}}\advancepageno 
  \fi}
\def\almostshipout#1{\if L\l@r \count1=1 \message{[\the\count0.\the\count1]}
      \global\setbox\leftpage=#1 \global\let\l@r=R
 \else \count1=2
  \shipout\vbox{\speclscape{\hsize\fullhsize\makeheadline}
      \hbox to\fullhsize{\box\leftpage\hfil#1}}  \global\let\l@r=L\fi}
\fi
%---------------------------------------------------------------------
%
\newcount\yearltd\yearltd=\year\advance\yearltd by -1900

\def\Title#1#2{\nopagenumbers\abstractfont\hsize=\hstitle\rightline{#1}%
\vskip 1in\centerline{\titlefont #2}\abstractfont\vskip .5in\pageno=0}
\def\Date#1{\vfill\leftline{#1}\tenpoint\supereject\global\hsize=\hsbody%
\footline={\hss\tenrm\folio\hss}}% 	restores pagenumbers
%
%       use following instead of \Date on the preliminary draft, 
%       puts date/time on each page in big mode, writes labels in margins

\def\draftmode{\message{ DRAFTMODE }\def\draftdate{{\rm preliminary draft:
\number\month/\number\day/\number\yearltd\ \ \hourmin}}%
\headline={\hfil\draftdate}\writelabels\baselineskip=20pt plus 2pt minus 2pt
 {\count255=\time\divide\count255 by 60 \xdef\hourmin{\number\count255}
  \multiply\count255 by-60\advance\count255 by\time
  \xdef\hourmin{\hourmin:\ifnum\count255<10 0\fi\the\count255}}}
%       use \nolabels to get rid of eqn, ref, and fig labels in draft mode
\def\nolabels{\def\wrlabeL##1{}\def\eqlabeL##1{}\def\reflabeL##1{}}
\def\writelabels{\def\wrlabeL##1{\leavevmode\vadjust{\rlap{\smash%
{\line{{\escapechar=` \hfill\rlap{\sevenrm\hskip.03in\string##1}}}}}}}%
\def\eqlabeL##1{{\escapechar-1\rlap{\sevenrm\hskip.05in\string##1}}}%
\def\reflabeL##1{\noexpand\llap{\noexpand\sevenrm\string\string\string##1}}}
\nolabels
%
% tagged sec numbers
\global\newcount\secno \global\secno=0
\global\newcount\meqno \global\meqno=1
\def\newsec#1{\global\advance\secno by1\message{(\the\secno. #1)}
%\ifx\answ\bigans \vfill\eject \else \bigbreak\bigskip \fi  %if desired
\global\subsecno=0\eqnres@t\noindent{\bf\the\secno. #1}
\writetoca{{\secsym} {#1}}\par\nobreak\medskip\nobreak}
\def\eqnres@t{\xdef\secsym{\the\secno.}\global\meqno=1\bigbreak\bigskip}
\def\sequentialequations{\def\eqnres@t{\bigbreak}}\xdef\secsym{}
\global\newcount\subsecno \global\subsecno=0
\def\subsec#1{\global\advance\subsecno by1\message{(\secsym\the\subsecno. #1)}
\ifnum\lastpenalty>9000\else\bigbreak\fi
\noindent{\it\secsym\the\subsecno. #1}\writetoca{\string\quad 
{\secsym\the\subsecno.} {#1}}\par\nobreak\medskip\nobreak}
\def\appendix#1#2{\global\meqno=1\global\subsecno=0\xdef\secsym{\hbox{#1.}}
\bigbreak\bigskip\noindent{\bf Appendix #1. #2}\message{(#1. #2)}
\writetoca{Appendix {#1.} {#2}}\par\nobreak\medskip\nobreak}
%
%       \eqn\label{a+b=c}	gives displayed equation, numbered
%				consecutively within sections.
%     \eqnn and \eqna define labels in advance (of eqalign?)
%
\def\eqnn#1{\xdef #1{(\secsym\the\meqno)}\writedef{#1\leftbracket#1}%
\global\advance\meqno by1\wrlabeL#1}
\def\eqna#1{\xdef #1##1{\hbox{$(\secsym\the\meqno##1)$}}
\writedef{#1\numbersign1\leftbracket#1{\numbersign1}}%
\global\advance\meqno by1\wrlabeL{#1$\{\}$}}
\def\eqn#1#2{\xdef #1{(\secsym\the\meqno)}\writedef{#1\leftbracket#1}%
\global\advance\meqno by1$$#2\eqno#1\eqlabeL#1$$}
%
%			 footnotes
\newskip\footskip\footskip14pt plus 1pt minus 1pt %sets footnote baselineskip
\def\footnotefont{\ninepoint}\def\f@t#1{\footnotefont #1\@foot}
\def\f@@t{\baselineskip\footskip\bgroup\footnotefont\aftergroup\@foot\let\next}
\setbox\strutbox=\hbox{\vrule height9.5pt depth4.5pt width0pt}
\global\newcount\ftno \global\ftno=0
\def\foot{\global\advance\ftno by1\footnote{$^{\the\ftno}$}}
%
%say \footend to put footnotes at end
%will cause problems if \ref used inside \foot, instead use \nref before
\newwrite\ftfile   
\def\footend{\def\foot{\global\advance\ftno by1\chardef\wfile=\ftfile
$^{\the\ftno}$\ifnum\ftno=1\immediate\openout\ftfile=foots.tmp\fi%
\immediate\write\ftfile{\noexpand\smallskip%
\noexpand\item{f\the\ftno:\ }\pctsign}\findarg}%
\def\footatend{\vfill\eject\immediate\closeout\ftfile{\parindent=20pt
\centerline{\bf Footnotes}\nobreak\bigskip\input foots.tmp }}}
\def\footatend{}
%
%     \ref\label{text}
% generates a number, assigns it to \label, generates an entry.
% To list the refs on a separate page,  \listrefs
%
\global\newcount\refno \global\refno=1
\newwrite\rfile
\def\ref{[\the\refno]\nref}
\def\nref#1{\xdef#1{[\the\refno]}\writedef{#1\leftbracket#1}%
\ifnum\refno=1\immediate\openout\rfile=refs.tmp\fi
\global\advance\refno by1\chardef\wfile=\rfile\immediate
\write\rfile{\noexpand\item{#1\ }\reflabeL{#1\hskip.31in}\pctsign}\findarg}
%	horrible hack to sidestep tex \write limitation
\def\findarg#1#{\begingroup\obeylines\newlinechar=`\^^M\pass@rg}
{\obeylines\gdef\pass@rg#1{\writ@line\relax #1^^M\hbox{}^^M}%
\gdef\writ@line#1^^M{\expandafter\toks0\expandafter{\striprel@x #1}%
\edef\next{\the\toks0}\ifx\next\em@rk\let\next=\endgroup\else\ifx\next\empty%
\else\immediate\write\wfile{\the\toks0}\fi\let\next=\writ@line\fi\next\relax}}
\def\striprel@x#1{} \def\em@rk{\hbox{}} 
\def\lref{\begingroup\obeylines\lr@f}
\def\lr@f#1#2{\gdef#1{\ref#1{#2}}\endgroup\unskip}

\def\addref#1{\immediate\write\rfile{\noexpand\item{}#1}} %now unnecessary
\def\startrefs#1{\immediate\openout\rfile=refs.tmp\refno=#1}
\def\xref{\expandafter\xr@f}\def\xr@f[#1]{#1}
\def\refs#1{\count255=1[\r@fs #1{\hbox{}}]}
\def\r@fs#1{\ifx\und@fined#1\message{reflabel \string#1 is undefined.}%
\nref#1{need to supply reference \string#1.}\fi%
\vphantom{\hphantom{#1}}\edef\next{#1}\ifx\next\em@rk\def\next{}%
\else\ifx\next#1\ifodd\count255\relax\xref#1\count255=0\fi%
\else#1\count255=1\fi\let\next=\r@fs\fi\next}
%

%
% this is ugly, but moore insists
\newwrite\ffile\global\newcount\figno \global\figno=1
\def\fig{fig.~\the\figno\nfig}
\def\nfig#1{\xdef#1{fig.~\the\figno}%
\writedef{#1\leftbracket fig.\noexpand~\the\figno}%
\ifnum\figno=1\immediate\openout\ffile=figs.tmp\fi\chardef\wfile=\ffile%
\immediate\write\ffile{\noexpand\medskip\noexpand\item{Fig.\ \the\figno. }
\reflabeL{#1\hskip.55in}\pctsign}\global\advance\figno by1\findarg}
\def\xfig{\expandafter\xf@g}\def\xf@g fig.\penalty\@M\ {}
\def\figs#1{figs.~\f@gs #1{\hbox{}}}
\def\f@gs#1{\edef\next{#1}\ifx\next\em@rk\def\next{}\else
\ifx\next#1\xfig #1\else#1\fi\let\next=\f@gs\fi\next}
\newwrite\lfile
{\escapechar-1\xdef\pctsign{\string\%}\xdef\leftbracket{\string\{}
\xdef\rightbracket{\string\}}\xdef\numbersign{\string\#}}

\def\writestop{\def\writestoppt{\immediate\write\lfile{\string\pageno%
\the\pageno\string\startrefs\leftbracket\the\refno\rightbracket%
\string\def\string\secsym\leftbracket\secsym\rightbracket%
\string\secno\the\secno\string\meqno\the\meqno}\immediate\closeout\lfile}}
\def\writestoppt{}\def\writedef#1{}
\def\seclab#1{\xdef #1{\the\secno}\writedef{#1\leftbracket#1}\wrlabeL{#1=#1}}
\def\subseclab#1{\xdef #1{\secsym\the\subsecno}%
\writedef{#1\leftbracket#1}\wrlabeL{#1=#1}}
\newwrite\tfile \def\writetoca#1{}
\def\leaderfill{\leaders\hbox to 1em{\hss.\hss}\hfill}
%	use this to write file with table of contents
\def\writetoc{\immediate\openout\tfile=toc.tmp 
   \def\writetoca##1{{\edef\next{\write\tfile{\noindent ##1 
   \string\leaderfill {\noexpand\number\pageno} \par}}\next}}}
%       and this lists table of contents on second pass

%
\catcode`\@=12 % at signs are no longer letters
%
%	Unpleasantness in calling in abstract and title fonts
\edef\tfontsize{\ifx\answ\bigans scaled\magstep3\else scaled\magstep4\fi}
\font\titlerm=cmr10 \tfontsize \font\titlerms=cmr7 \tfontsize
\font\titlermss=cmr5 \tfontsize \font\titlei=cmmi10 \tfontsize
\font\titleis=cmmi7 \tfontsize \font\titleiss=cmmi5 \tfontsize
\font\titlesy=cmsy10 \tfontsize \font\titlesys=cmsy7 \tfontsize
\font\titlesyss=cmsy5 \tfontsize \font\titleit=cmti10 \tfontsize
\skewchar\titlei='177 \skewchar\titleis='177 \skewchar\titleiss='177
\skewchar\titlesy='60 \skewchar\titlesys='60 \skewchar\titlesyss='60
\def\titlefont{\def\rm{\fam0\titlerm}% switch to title font
\textfont0=\titlerm \scriptfont0=\titlerms \scriptscriptfont0=\titlermss
\textfont1=\titlei \scriptfont1=\titleis \scriptscriptfont1=\titleiss
\textfont2=\titlesy \scriptfont2=\titlesys \scriptscriptfont2=\titlesyss
\textfont\itfam=\titleit \def\it{\fam\itfam\titleit}\rm}
 \ifx\answ\bigans\else scaled\magstep1\fi
\ifx\answ\bigans\def\abstractfont{\tenpoint}\else
\font\abssl=cmsl10 scaled \magstep1
\font\absrm=cmr10 scaled\magstep1 \font\absrms=cmr7 scaled\magstep1
\font\absrmss=cmr5 scaled\magstep1 \font\absi=cmmi10 scaled\magstep1
\font\absis=cmmi7 scaled\magstep1 \font\absiss=cmmi5 scaled\magstep1
\font\abssy=cmsy10 scaled\magstep1 \font\abssys=cmsy7 scaled\magstep1
\font\abssyss=cmsy5 scaled\magstep1 \font\absbf=cmbx10 scaled\magstep1
\skewchar\absi='177 \skewchar\absis='177 \skewchar\absiss='177
\skewchar\abssy='60 \skewchar\abssys='60 \skewchar\abssyss='60
\def\abstractfont{\def\rm{\fam0\absrm}% switch to abstract font
\textfont0=\absrm \scriptfont0=\absrms \scriptscriptfont0=\absrmss
\textfont1=\absi \scriptfont1=\absis \scriptscriptfont1=\absiss
\textfont2=\abssy \scriptfont2=\abssys \scriptscriptfont2=\abssyss
\textfont\itfam=\bigit \def\it{\fam\itfam\bigit}\def\footnotefont{\tenpoint}%
\textfont\slfam=\abssl \def\sl{\fam\slfam\abssl}%
\textfont\bffam=\absbf \def\bf{\fam\bffam\absbf}\rm}\fi
\def\tenpoint{\def\rm{\fam0\tenrm}% switch back to 10-point type
\textfont0=\tenrm \scriptfont0=\sevenrm \scriptscriptfont0=\fiverm
\textfont1=\teni  \scriptfont1=\seveni  \scriptscriptfont1=\fivei
\textfont2=\tensy \scriptfont2=\sevensy \scriptscriptfont2=\fivesy
\textfont\itfam=\tenit \def\it{\fam\itfam\tenit}\def\footnotefont{\ninepoint}%
\textfont\bffam=\tenbf \def\bf{\fam\bffam\tenbf}\def\sl{\fam\slfam\tensl}\rm}
\font\ninerm=cmr9 \font\sixrm=cmr6 \font\ninei=cmmi9 \font\sixi=cmmi6 
\font\ninesy=cmsy9 \font\sixsy=cmsy6 \font\ninebf=cmbx9 
\font\nineit=cmti9 \font\ninesl=cmsl9 \skewchar\ninei='177
\skewchar\sixi='177 \skewchar\ninesy='60 \skewchar\sixsy='60 
\def\ninepoint{\def\rm{\fam0\ninerm}% switch to footnote font
\textfont0=\ninerm \scriptfont0=\sixrm \scriptscriptfont0=\fiverm
\textfont1=\ninei \scriptfont1=\sixi \scriptscriptfont1=\fivei
\textfont2=\ninesy \scriptfont2=\sixsy \scriptscriptfont2=\fivesy
\textfont\itfam=\ninei \def\it{\fam\itfam\nineit}\def\sl{\fam\slfam\ninesl}%
\textfont\bffam=\ninebf \def\bf{\fam\bffam\ninebf}\rm} 
%
%---------------------------------------------------------------------
%

\hyphenation{anom-aly anom-alies coun-ter-term coun-ter-terms}
\def\inv{^{\raise.15ex\hbox{${\scriptscriptstyle -}$}\kern-.05em 1}}

\def\Dsl{\,\raise.15ex\hbox{/}\mkern-13.5mu D} %this one can be subscripted
\def\dsl{\raise.15ex\hbox{/}\kern-.57em\partial}

\font\bigit=cmti10 scaled \magstep1
 %pound sterling
\def\lspace{\ifx\answ\bigans{}\else\qquad\fi}
\def\lbspace{\ifx\answ\bigans{}\else\hskip-.2in\fi} % $$\lbspace...$$
\def\boxeqn#1{\vcenter{\vbox{\hrule\hbox{\vrule\kern3pt\vbox{\kern3pt
	\hbox{${\displaystyle #1}$}\kern3pt}\kern3pt\vrule}\hrule}}}
\def\mbox#1#2{\vcenter{\hrule \hbox{\vrule height#2in
		\kern#1in \vrule} \hrule}}  %e.g. \mbox{.1}{.1}
%	matters of taste
%\def\tilde{\widetilde} \def\bar{\overline} \def\hat{\widehat}
%
% some sample definitions
  %     curly letters

\def\darr#1{\raise1.5ex\hbox{$\leftrightarrow$}\mkern-16.5mu #1}
 %pound sterling

 %puts a small half in a displayed eqn
\def\roughly#1{\raise.3ex\hbox{$#1$\kern-.75em\lower1ex\hbox{$\sim$}}}

%\draftmode
\let\includefigures=\iftrue
\let\useblackboard=\iftrue
\newfam\black

%Figure Stuff
\includefigures
\message{If you do not have epsf.tex (to include figures),}
\message{change the option at the top of the tex file.}
\input epsf
\def\figin{\epsfcheck\figin}\def\figins{\epsfcheck\figins}
\def\epsfcheck{\ifx\epsfbox\UnDeFiNeD
\message{(NO epsf.tex, FIGURES WILL BE IGNORED)}
\gdef\figin##1{\vskip2in}\gdef\figins##1{\hskip.5in}% blank space instead
\else\message{(FIGURES WILL BE INCLUDED)}%
\gdef\figin##1{##1}\gdef\figins##1{##1}\fi}
\def\DefWarn#1{}
\def\figinsert{\goodbreak\midinsert}
\def\ifig#1#2#3{\DefWarn#1\xdef#1{fig.~\the\figno}
\writedef{#1\leftbracket fig.\noexpand~\the\figno}%
\figinsert\figin{\centerline{#3}}\medskip\centerline{\vbox{
\baselineskip12pt\advance\hsize by -1truein
\noindent\footnotefont{\bf Fig.~\the\figno:} #2}}
%\bigskip
\endinsert\global\advance\figno by1}
%%%
\else
\def\ifig#1#2#3{\xdef#1{fig.~\the\figno}
\writedef{#1\leftbracket fig.\noexpand~\the\figno}%
%\figinsert\figin{\centerline{#3}}\medskip
%\centerline{\vbox{\baselineskip12pt
%\advance\hsize by -1truein\noindent
%\footnotefont{\bf Fig.~\the\figno:} #2}}
%\bigskip\endinsert
\global\advance\figno by1} \fi

\def\id{{1 \kern-.28em {\rm l}}}

\def\K3{{\bf K3}}
\def\journal#1&#2(#3){\unskip, \sl #1\ \bf #2 \rm(19#3) }
\def\andjournal#1&#2(#3){\sl #1~\bf #2 \rm (19#3) }

\def\bar{\overline}

\def\tilde{\widetilde}

\def\frac#1#2{{#1\over#2}}

\def\inbar{\,\vrule height1.5ex width.4pt depth0pt}
\def\IC{\relax\hbox{$\inbar\kern-.3em{\rm C}$}}
\def\IR{\relax{\rm I\kern-.18em R}}
\def\IP{\relax{\rm I\kern-.18em P}}

%
%%%%%%%%%%%%%%%%%%%%%%%%%%%%%%%%%%%%
%

%
\catcode`\@=11
\def\slash#1{\mathord{\mathpalette\c@ncel{#1}}}
\overfullrule=0pt

\def\NN{{\cal N}}
\def\OO{{\cal O}}

\def\underrel#1\over#2{\mathrel{\mathop{\kern\z@#1}\limits_{#2}}}

\catcode`\@=12

%%%%%%%%%%%%%%%%%%%%%%%%%%%%%%%%%%%%%%%%%%%%%%%%%%%%%%%%%%%%%%

%

%%%%%%%%%%%%%%%%%%%%%%%%%%%%%%%%%%%%%%%%%%%%%%%%%%%%%%%%%%%%%%
% new defs:

\def\p{{\partial}}

\def\ra{{\rightarrow}}

\def\tc{{\tilde c}}

\def\tmu{{\tilde \mu}}
\def\tt{{\tilde t}}
\def\dS{{\delta S}}

%\SakaiCN
\lref\SakaiCN{
T.~Sakai and S.~Sugimoto,
``Low energy hadron physics in holographic QCD,''
Prog.\ Theor.\ Phys.\  {\bf 113}, 843 (2005)
[arXiv:hep-th/0412141].
%%CITATION = HEP-TH 0412141;%%
%%Cited 33 times in SPIRES-HEP
}
\lref\AHJK{
  E.~Antonyan, J.~A.~Harvey, S.~Jensen and D.~Kutasov,
  ``NJL and QCD from string theory,''
  arXiv:hep-th/0604017.
  %%CITATION = HEP-TH 0604017;%%
}

%\WittenZW
\lref\WittenZW{
  E.~Witten,
  ``Anti-de Sitter space, thermal phase transition, and confinement in  gauge
  theories,''
  Adv.\ Theor.\ Math.\ Phys.\  {\bf 2}, 505 (1998)
  [arXiv:hep-th/9803131].
  %%CITATION = 00203,2,505;%%
}

%\AharonyDA
\lref\AharonyDA{
  O.~Aharony, J.~Sonnenschein and S.~Yankielowicz,
  ``A holographic model of deconfinement and chiral symmetry restoration,''
  Annals Phys.\  {\bf 322}, 1420 (2007)
  [arXiv:hep-th/0604161].
  %%CITATION = APNYA,322,1420;%%
}

%\ParnachevDN
\lref\ParnachevDN{
  A.~Parnachev and D.~A.~Sahakyan,
  ``Chiral phase transition from string theory,''
  Phys.\ Rev.\ Lett.\  {\bf 97}, 111601 (2006)
  [arXiv:hep-th/0604173].
  %%CITATION = PRLTA,97,111601;%%
}

%\ParnachevEV
\lref\ParnachevEV{
  A.~Parnachev and D.~A.~Sahakyan,
  ``Photoemission with chemical potential from QCD gravity dual,''
  Nucl.\ Phys.\  B {\bf 768}, 177 (2007)
  [arXiv:hep-th/0610247].
  %%CITATION = NUPHA,B768,177;%%
}

%\HorigomeXU
\lref\HorigomeXU{
  N.~Horigome and Y.~Tanii,
  ``Holographic chiral phase transition with chemical potential,''
  JHEP {\bf 0701}, 072 (2007)
  [arXiv:hep-th/0608198].
  %%CITATION = JHEPA,0701,072;%%
}

%\KimGP
\lref\KimGP{
  K.~Y.~Kim, S.~J.~Sin and I.~Zahed,
  ``Dense hadronic matter in holographic QCD,''
  arXiv:hep-th/0608046.
  %%CITATION = HEP-TH/0608046;%%
}

%\KimZM
\lref\KimZM{
  K.~Y.~Kim, S.~J.~Sin and I.~Zahed,
  ``The Chiral Model of Sakai-Sugimoto at Finite Baryon Density,''
  arXiv:0708.1469 [hep-th].
  %%CITATION = ARXIV:0708.1469;%%
}

%\BergmanWP
\lref\BergmanWP{
  O.~Bergman, G.~Lifschytz and M.~Lippert,
  ``Holographic Nuclear Physics,''
  arXiv:0708.0326 [hep-th].
  %%CITATION = ARXIV:0708.0326;%%
}

%\DavisKA
\lref\DavisKA{
  J.~L.~Davis, M.~Gutperle, P.~Kraus and I.~Sachs,
  ``Stringy NJL and Gross-Neveu models at finite density and temperature,''
  arXiv:0708.0589 [hep-th].
  %%CITATION = ARXIV:0708.0589;%%
}

%\RozaliRX
\lref\RozaliRX{
  M.~Rozali, H.~H.~Shieh, M.~Van Raamsdonk and J.~Wu,
  ``Cold Nuclear Matter In Holographic QCD,''
  arXiv:0708.1322 [hep-th].
  %%CITATION = ARXIV:0708.1322;%%
}

%\SonXC
\lref\SonXC{
  D.~T.~Son and M.~A.~Stephanov,
  ``QCD at finite isospin density,''
  Phys.\ Rev.\ Lett.\  {\bf 86}, 592 (2001)
  [arXiv:hep-ph/0005225].
  %%CITATION = PRLTA,86,592;%%
}

%\KleinFY
\lref\KleinFY{
  B.~Klein, D.~Toublan and J.~J.~M.~Verbaarschot,
  ``The QCD phase diagram at nonzero temperature, baryon and isospin  chemical
  potentials in random matrix theory,''
  Phys.\ Rev.\  D {\bf 68}, 014009 (2003)
  [arXiv:hep-ph/0301143].
  %%CITATION = PHRVA,D68,014009;%%
}

%\ToublanTT
\lref\ToublanTT{
  D.~Toublan and J.~B.~Kogut,
  ``Isospin chemical potential and the QCD phase diagram at nonzero
  temperature and baryon chemical potential,''
  Phys.\ Lett.\  B {\bf 564}, 212 (2003)
  [arXiv:hep-ph/0301183].
  %%CITATION = PHLTA,B564,212;%%
}

%\BarducciTT
\lref\BarducciTT{
  A.~Barducci, R.~Casalbuoni, G.~Pettini and L.~Ravagli,
  ``A calculation of the QCD phase diagram at finite temperature, and  baryon
  and isospin chemical potentials,''
  Phys.\ Rev.\  D {\bf 69}, 096004 (2004)
  [arXiv:hep-ph/0402104].
  %%CITATION = PHRVA,D69,096004;%%
}

%\ErdmengerAP
\lref\ErdmengerAP{
  J.~Erdmenger, M.~Kaminski and F.~Rust,
  ``Isospin diffusion in thermal AdS/CFT with flavor,''
  Phys.\ Rev.\  D {\bf 76}, 046001 (2007)
  [arXiv:0704.1290 [hep-th]].
  %%CITATION = PHRVA,D76,046001;%%
}

%\AntonyanQY
\lref\AntonyanQY{
  E.~Antonyan, J.~A.~Harvey and D.~Kutasov,
  ``The Gross-Neveu model from string theory,''
  Nucl.\ Phys.\  B {\bf 776}, 93 (2007)
  [arXiv:hep-th/0608149].
  %%CITATION = NUPHA,B776,93;%%
}

%\AntonyanPG
\lref\AntonyanPG{
  E.~Antonyan, J.~A.~Harvey and D.~Kutasov,
  ``Chiral symmetry breaking from intersecting D-branes,''
  Nucl.\ Phys.\ Proc.\ Suppl.\  {\bf 171}, 243 (2007)
  [arXiv:hep-th/0608177].
  %%CITATION = NUPHZ,171,243;%%
}

\Title{\vbox{\baselineskip12pt
%\hbox{hep-th/yymmnnn}
}}
{\vbox{\centerline{Holographic QCD with Isospin Chemical Potential}
%    \centerline{with Isospin Chemical Potential}
\vskip.06in
}}
\centerline{Andrei Parnachev}
\bigskip
\centerline{{\it C.N.Yang Institute for Theoretical Physics, SUNY, Stony Brook, NY 11794-3840, USA}}
\vskip.1in \vskip.1in \centerline{\bf Abstract}  
\noindent
We consider configurations of $D4-D8-\bar D8$ branes which correspond to large $N$ 
QCD with non-vanishing temperature and chemical potential for
baryon number and isospin.
We study the holographic dual of this model and find a rich phase structure.
The phases, distinguished by the values of quark condensates,
are separated by the surfaces of first order phase transitions.
The picture is in many respects similar to the expected phase
structure of QCD in the chiral limit.

\vfill

\Date{August 2007}
   
%\draftmode

%%%%%%%%%%%%%%%%%%%%%%%%%%%%%%%%%%%%%%%%%%%%%%
\newsec{Introduction and summary}

String theoretic constructions aimed at describing the theory of
strong interactions, QCD, are becoming more and more elaborate.
There are now many versions of the gauge/string correspondence 
where many constraints of the $\NN=4$ Super Yang Mills are relaxed,
although one still has to consider the limit of large number of colors $N$. 
However the description of large $N$ QCD-like (asymptotically free) theories
proves to be a hard task.
Typically, the regime where the gauge theory is modified at energies
much higher than the dynamically generated scale, is not easily tractable
on the dual, string theoretic  side.

Despite these limitations, holographic models are still useful
because very often they capture physics qualitatively correct.
Hence, they provide a unique window into the dynamics of strongly coupled
gauge theories.
With more luck, it might be possible to extract quantities that
are universal, i.e. do not significantly change as one interpolates 
between the holographic dual and QCD.
A good place to search for such universal quantities is the phase 
diagram of the theory, especially if the second order phase 
transitions are present.

A model proposed in \SakaiCN\ building on \WittenZW\ which 
involves a certain configuration of  $D4-D8-\bar D8$ branes is
particularly interesting because it gives rise to large $N$ QCD
with quarks in a certain region of the parameter space.
This model, also sometimes called ``holographic QCD'', 
has a holographic dual which reproduces some expected physics
of the gauge theory.
Chiral symmetry breaking and  derivation of the pion lagrangian have been
described in the original paper by Sakai and Sugimoto \SakaiCN.
The model has been also investigated at finite temperature \refs{\AharonyDA,\ParnachevDN}.
Nonzero baryon chemical potential has been considered in \refs{\HorigomeXU,\ParnachevEV}
where the phase diagram was shown to contain a line of first order phase transitions 
in the $\mu_B-T$ plane.\foot{In the first versions of \HorigomeXU\ the 
solution with broken chiral symmetry contained an unphysical
charge. This has been corrected in  \ParnachevEV\ and in the most recent
version of \HorigomeXU.
Recent work which discusses related physics and possible additional
phases includes  \refs{\KimGP\KimZM\BergmanWP\DavisKA-\RozaliRX}.
The isospin chemical potential in the $D3-D7$ system was recently 
discussed in  \ErdmengerAP. }

In this paper we focus our attention on the case of two flavor ($N_f=2$) large
$N$ holographic QCD with non-vanishing isospin chemical potential, whose 
value we denote by $\mu_I$.
QCD with $\mu_I\neq0$ has been studied in the variety of models
(see \refs{\SonXC\KleinFY\ToublanTT-\BarducciTT} for an incomplete
list of references.)
The phase diagram enjoys rich structure, with first and second
order phase transition lines separating various phases.
This therefore is a promising setting for the search
of universal quantities.
Besides, theory with non-vanishing $\mu_I$ is amenable to
lattice simulations and does not suffer from the determinant sign
problem, which complicates obtaining lattice results for the $\mu_B\neq0$ case.
Finally, non-vanishing isospin chemical potential can be experimentally
relevant.

The holographic dual of the  $D4-D8-\overline{D8}$ branes involves
the warped product of four-dimensional Minkowski space, a circle
parameterized by $X^4\in[0,2\pi R_4)$, a holographic coordinate $U>U_K$
and a four-sphere (more details are provided below).
At non-zero temperature the phase where
gluons are not confined is described by a black hole metric
which is thermodynamically preferred for temperatures higher than $1/2\pi R_4$.
Most of the analysis below will be restricted to this phase.
We comment on the appearance of confinement/deconfinement transition in 
the phase diagram in Section 4.

Fundamental matter is described by $N_f$ pairs of  $D8-\overline{D8}$ branes
with asymptotic separation $L$.
The low energy degrees of freedom which live on the intersections
of the $D8$ and $\overline{D8}$ branes with the $D4$ branes are
four-dimensional massless left and right-handed quarks.
This brane setup thus describes large $N$ QCD in the chiral 
limit (in the certain region of the parameter space).
At generic values of baryon and isospin chemical potentials, the
$D8$ branes exist in two phases. In the curved phase $D8$ and $\overline{D8}$
branes connect and hence chiral symmetry is broken.
In the straight phase, the flavor branes are located at constant 
values of $X^4$ and fall into the horizon of the black hole.
This phase of restored chiral symmetry is preferred at larger values of
$T$ and $\mu_B$.

The phase diagram of the system depends on dimensionless parameters $LT$ and $\tmu_i=\mu_i/E_0$
where $E_0=4\pi\lambda/L^2$ sets the scale of chemical potential.
($\lambda$ is the t'Hooft coupling of five-dimensional gauge theory
on the $D4$ branes, which becomes four-dimensional after Kaluza-Klein reduction.)
Fig. 1 contains the picture of the phase diagram in the limit $LT\ra0$.
(It is instructive to compare it with Fig. 1 in \KleinFY.)
The fundamental domain can be taken to be 
\eqn\fdomain{ -\tmu_1\le\tmu_2\le\tmu_1;\;\tmu_1\ge0  }
The phases are distinguished by the values of the condensates:
\eqn\conds{  \sigma_i=\langle\bar\psi_i\psi_i\rangle, \;i=1,2;\qquad
             \rho=\langle\bar\psi_i\gamma^5\tau^2_{ij}\psi_j\rangle  }
where $\tau^3$ is the Pauli matrix in the flavor space.
In the upper right triangle defined by $\tmu_1,\tmu_2\ge\tmu_c=0.11$
all condensates vanish.
Curved blue line separates the phase with non-zero pion condensate $\rho$ at
small $\mu_B$ (below the line) from the phases with $\sigma_1\neq0,\;\sigma_2=0$ 
for $-\tmu_c\le\tmu_2\le\tmu_c$ and $\sigma_{1,2}=0$ for $\tmu_2\le\tmu_c$.
All these features are also present in Fig. 1 of ref. \KleinFY.
What is not present in our analysis is the second order phase transition
of \KleinFY\ which separates the phase with $\rho\neq0$ at small $\tmu_I$
from the phase with $\sigma_{1,2},\rho=0$ at larger $\tmu_I$.
This line might be an artifact of the random matrix model used
in \KleinFY.\foot{In the view of our motivation, the absence of the second order
phase transition is a slight disappointment.}
%It is interesting to learn, however, that the existence of this second order line
%is not a universal property.

So far we discussed the $LT\ra0$ case. The diagram does not qualitatively
change as we increase $LT$ (see Figs. 5,6), although the value of $\tmu_c$ decreases
until it reaches zero at $LT\approx0.15$ (Fig. 4).
This corresponds to the temperature of chiral phase transition
at $\mu_B=0$.
The full phase diagram should be drawn in three dimensions, 
with Figs. 1,2,5,6 representing slices at two different values
of $LT$.
The surface of the first order phase transitions which is
represented by the blue curves on the pictures of these slices
looks in fact somewhat similar to a hyperboloid, with $\mu_I$ 
axis being the axis of rotation.
(Of course, there is no rotational symmetry in the $\mu_B-T$ plane
in our case)

The rest of the paper is organized as follows.
In the next section we set up the notation, 
and discuss general expectations.
We then consider  $LT\ra0$ case, establish the necessary holographic
dictionary and present some analytic and numerical results.
In Section 3 we consider finite temperature case, where most
of the analysis is numerical. Here we sketch the form of
the three-dimensional phase diagram.
We conclude in Section 4.

%%%%%%%%%%%%%%%%%%%%%%%%%%%%%%%%%%%%%%%%%%%%%%%%%%%
\newsec{Nonzero chemical potentials; small temperature}

We consider $SU(N)$ gauge theory with $N_f=2$ massless quarks (each described by a single
Dirac spinor $\psi_i,\; i=1,2$) in the large $N$ limit.
Various phases of the theory can be distinguished by the
values of the condensates \conds.
The matter part of the Lagrangian is
\eqn\mlag{ {\cal L}_q= \sum_{i=1,2} {\bar \psi}_i 
        [\gamma^\mu (\p_\mu+i A_\mu) +\mu_i\gamma^0]\psi_i}
where sum over color indexes of $SU(N)$ is implied.
Baryon and isospin chemical potential are defined to be
\eqn\bacp{  \mu_B={1\over2}(\mu_1+\mu_2);\qquad \mu_I={1\over2}(\mu_1-\mu_2) }
respectively.
At $\mu_{1,2}=0$ the Lagrangian has $SU(2)\times SU(2)$ global
symmetry.
One can use the axial part of this symmetry to rotate the $\sigma_{1,2}$
condensates into $\rho$; these condensates develop nonzero expectation
values, and spontaneously break chiral symmetry.
The corresponding Nambu-Goldstone model is the pion.
At non-zero value of $\mu_I$ (and sufficiently small $\mu_{1,2}$)
the system is supposed to prefer nonzero value for $\rho$ together
with $\sigma_{1,2}=0$. (see e.g. \KleinFY.)

This large $N$ QCD has a holographic dual, which is obtained
by going to the near-horizon geometry of the $D4$ branes on a circle
of radius $R_4$ \WittenZW\ and
adding $N_f=2$ $D8-\bar D8$ pairs \SakaiCN.
The model deviates from QCD at energy scales $\sim 1/R_4$
which is related to the dynamically generated scale as
\eqn\lqcd{  \Lambda_{QCD}\sim {1\over R_4} e^{-{1\over\lambda_4}}  }
where $\lambda_4=\lambda/R_4$ is the four-dimensional t'Hooft
coupling at the cutoff scale.
As customary in the holographic models, when the scales are widely separated,
$\lambda_4$ is small, which corresponds to the strong curvature
regime in string theory.
This is beyond the reach of the present string-theoretic technology.
We consider instead the opposite regime of large $\lambda_4$ where
string theory reduces to supergravity + low energy $D8$-brane dynamics.

In this Section we restrict to the regime of small temperatures,  $LT\ra0$.
We assume that confinement energy scale 
is much smaller than the scale associated with the dynamics of
fundamental matter ($1/R_4\ll 1/L$), and even at these small temperature
glue is deconfined.\foot{Formally, this limit is equivalent to the NJL limit
studied in \refs{\AHJK\AntonyanQY-\AntonyanPG}}
(It is not hard to reinstate the effects of confinement/deconfinement
transition on the phase diagram - we do it in Section 4. )

The euclidean metric of the holographic dual is given by
\eqn\metric{  ds^2=\left({U\over R}\right)^{3\over2}(\eta_{\mu\nu} dX^\mu dX^\nu+(dX^4)^2)
      +\left({U\over R}\right)^{-{3\over2}}[(dU)^2+U^2d\Omega_4^2] }
There is also non-trivial dilaton 
\eqn\dilaton{  e^{\Phi}=g_s\left({ U\over R}\right)^{3\over4}  }
The length scale $R$ is related to the parameters of  the gauge theory as
\eqn\param{   R^3=\pi\lambda=\pi g_s N    }
Here and in the rest of the paper the string length $l_s=1$.
The dynamics of the fundamental matter is described by the DBI
action of $N_f=2$ $D8-\bar D8$ pairs. 
Asymptotically (as $U\ra\infty$) the branes and antibranes are separated
by coordinate distance $L$ in the $X^4$ direction.
They may connect at smaller values of $U$.
This configuration spontaneously breaks chiral symmetry.
Gauge fields on the $D8$ branes are holographically dual to
the currents of global  $SU(2)\times SU(2)$ symmetry.
As in \ParnachevEV\ it is convenient to go to the gauge $A_U=0$ and
Wick rotate $A_0\ra iA_0$.
Then, the non-zero chemical potential in the lagrangian \mlag\
corresponds to non-trivial boundary conditions on the
values of $A_0$:
\eqn\bcanot{   A_{0i} (U\ra\infty)=\mu_i;\qquad i=1,2    }
where $A_{0i}$ is the abelian part of the field on the
$i$-th brane and antibrane.

In the absence of external charges, baryon chemical 
potential cannot affect the curved solution \ParnachevEV.
Hence, for the purpose of computing the action, we can
set $\mu_1=-\mu_2=\mu_I$ in this phase.
There are now two possible curved brane solutions.
One solution involves the brane with $A_0(U\ra\infty)=\pm\mu$ connecting
with the antibrane with $A_0(U\ra\infty)=\pm\mu$.
This solution has zero electric field on the resulting curved flavor branes
and non-zero condensates $\sigma_1=\sigma_2$.
As explained in \AHJK\ the holographic dual of the chiral condensate
involves a non-trivial profile of the open string tachyon and hence
is difficult to analyze in the DBI approximation. 
% the branes 
%completely overlap and $SU(2)_V$ is unbroken.

Another solution involves brane with  $A_0(U\ra\infty)=\pm\mu$ connecting
with the antibrane with $A_0(U\ra\infty)=\mp\mu$.
There is nonzero electric field of opposite direction, but same magnitude 
on two curved branes so the branes overlap again.
This solution is obtained from the previous one by chiral 
rotation, and hence corresponds to nonzero value of $\rho$.
%In fact, the two solutions are obtained
%from each other by chiral rotation.
%The exact 
%This is the holographic dual of the statements made in the
%beginning of this section.

In the following we will use the fact that in the connected
phase $(\p_U A_{01})^2=(\p_U A_{02})^2$ and $X^4_1=X^4_2$
where the subscript corresponds to the isospin.
The DBI action, up to an overall constant, is given by
\eqn\dbia{  S=2\int dU U^{5\over2}\sqrt{1-(\p_U A_0)^2+
             \left({U\over R}\right)^3 (\p_U X^4)^2}  }
where a factor of two in front of the action counts the number of flavors
and is added for convenience.
(Since the only non-vanishing component of the gauge field is $A_0(U)$, the
Chern-Simons term vanishes.)
Lagrange-Euler equation of motion for the gauge field on the branes 
implies the existence of the conserved quantity,
\eqn\ccc{  {U^{5\over2} (\p_U A_0)\over\sqrt{1-(\p_U A_0)^2+
             \left({U\over R}\right)^3 (\p_U X^4)^2}}=c }
Eq. \ccc\ can also be written as
\eqn\eomgfb{  (\p_U A_0)^2=c^2\; {1+\left({U\over R}\right)^3 (\p_U X^4)^2\over U^5+
     c^2}}
%
%where $c$ is the conserved quantity.
Likewise, the equation of motion for $X^4$ can be written as
\eqn\eomtaua{ (\p_UX^4)^2={U_0^3 (U_0^5+c^2) R^3\over U^6 (U^5+c^2)-U_0^3 (U_0^5+c^2) U^3}
} 
where $U_0$ labels the turning point: it is the value of the $U$ 
coordinate where $\p_UX^4\ra\infty$.
To obtain \eomtaua\ we expressed $(\p_UX^4)$ in terms of
the constants of the equations of motion and then demanded 
$\p_UX^4\ra\infty$ at $U=U_0$.
Substituting \eomtaua\ back into \eomgfb\ gives
\eqn\eomgfc{ (\p_U A_0)^2= {c^2  U^3\over U^3 (U^5+U_0^5\tc^2)-U_0^8(1+\tc^2)}  }
where we introduced $\tc$ via $c^2=U_0^5 \tc^2$.
%
%Let us neglect the baryonic effects for now.
Let us consider the phase with $\p_U A_1=-\p_U A_2$.
%We are going to describe it below.
We need to re-express the values of the integrals of motion
through $\mu_{1,2}$ and $L$.
Starting with the former, we have
\eqn\mudiff{  {1\over2}|\mu_1-\mu_2|=\int_{U_0}^\infty dU |\p_U A_0|=U_0 F(\tc)  }
where we used \eomgfc\ in the second equality.
We also pass to the rescaled quantities and define $F(\tc)$ as
\eqn\fdef{ F(\tc)= \int_1^\infty dx \sqrt{x^3 \tc^2\over x^3(x^5+\tc^2)-(1+\tc^2)}  }  
Note that $F(\tc)$ is a monotonic function which satisfies $F(0)=0$ and
\eqn\fasysm{ F(\tc)\sim \tc +o(\tc),\qquad \tc\ll1}
In the opposite regime of large $\tc$
\eqn\fasyla{ F(\tc) \approx c_1 \tc^{2\over5},\qquad 
        c_1=\int_0^\infty {dx\over\sqrt{1+x^5} }={\Gamma\left({3\over10}\right)
                                               \Gamma\left({6\over5}\right)\over\sqrt{\pi}}    }
Similarly, the separation of the branes can be computed as
\eqn\lhalf{  {L\over2}=\int_{U_0}^\infty dU \p_U X^4=\sqrt{\pi\lambda\over U_0} G(\tc)  }
where 
\eqn\gdef{ G(\tc)=\int_1^\infty dx\sqrt{1+\tc^2\over x^6 (x^5+\tc^2)-x^3 (1+\tc^2)}  }
which is a monotonically increasing function with
\eqn\gzero{  G(0)= { 2\sqrt{\pi} \Gamma\left({9\over16}\right)\over\Gamma\left({1\over16}\right)},
\qquad   \lim_{\tc\ra\infty} G(\tc) = 
      { 2\sqrt{\pi}\; \Gamma\left({2\over3}\right)\over\Gamma\left({1\over6}\right)}}
We can use \lhalf\ to re-express 
\eqn\unotre{  U_0=E_0 G^2(\tc),\qquad E_0={4\pi\lambda\over L^2   }}
and then substitute the result into eq. \mudiff\
\eqn\mudiffa{  {1\over2}|\tmu_1-\tmu_2|= F(\tc) G^2(\tc)  }
where $\tmu\equiv\mu/E_0$.
Eq. \mudiffa\ determines the value of $\tc$ in terms of $\tmu$.
Next we need to compute the action.
To regularize, we subtract the value of the action from that of a pair 
of straight branes ($\p_U X^4=0$) with vanishing chemical potential.\foot{
Another popular regularization scheme involves introducing an ultraviolet (large $U$) cutoff
for the integral. Since we are interested in the difference between the actions for 
different configurations, these regularizations are equivalent.}
The resulting quantity is
\eqn\action{  \dS= 2 \left[\int_0^{U_0} dU U^{5\over2}+\int_{U_0}^\infty\left(
         U^{5\over2}-U^{5\over2}\sqrt{1-(\p_UA_0)^2+\left({U\over R}\right)^3 (\p_U x^4)^2}\right)\right]}
With the help of \unotre\ this can also be written as
\eqn\actiona{   \dS=2 E_0^{7\over2} G^7(\tc) H(\tc)   }
where 
\eqn\hdef{  H(\tc)={2\over7}+\int_1^\infty x^{5\over2}\;\left(1-\sqrt{x^8\over x^3(x^5+\tc^2)-(1+\tc^2)}\right)}
This should be compared with the action for the phase where the
branes are curved but have zero electric field.
(This is the phase with $\sigma_{1,2}\neq0,\rho=0$.)
\eqn\acvac{   \dS_0=2 E_0^{7\over2} G^7(0) H(0)   }
The thermodynamically preferred phase has a larger value of $\dS$.
One can check that \actiona\ is larger or equal to \acvac,
and hence as soon as $\mu_I\neq0$ the phase with $\rho\neq0$ is thermodynamically
preferred. 
An interesting limit of \hdef\ is $\tc\ra\infty$, which 
corresponds to the large values of $\tmu_I$,
\eqn\hla{  H(\tc) \approx {2 c_1\over7} \; \tc^{7\over5} }
which corresponds to
\eqn\sla{  \dS\approx {2 E_0^{7\over2}\over7 c_1^{5\over2} }\; {|\tmu_1-\tmu_2|^{7\over2}\over2^{5\over2}}}  
\midinsert\bigskip{\vbox{{\epsfxsize=3in
        \nobreak
    \centerline{\epsfbox{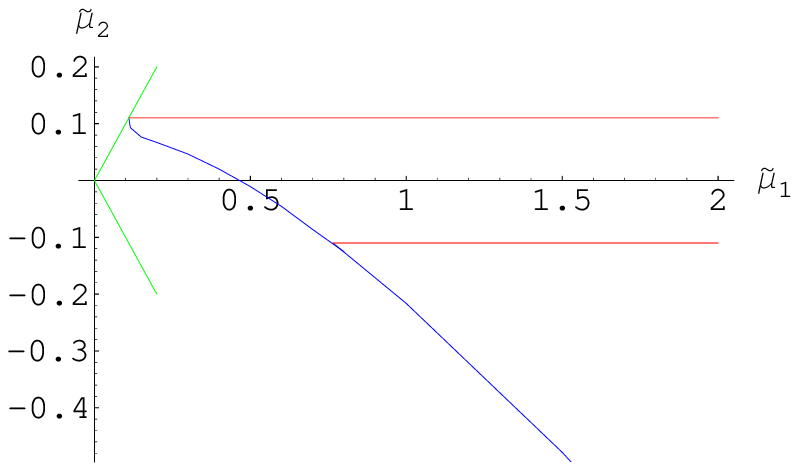}}
        \nobreak\bigskip
    {\raggedright\it \vbox{
{\bf Fig 1.}
{\it  Phase diagram for $LT\ra0$. $\mu_{1,2}$ are horizontal and vertical axes.
Green lines $\tmu_2=\pm\tmu_1$ are boundaries
of the fundamental domain. Solid blue and red curves are the lines of
first order phase transitions.
The phase between the red lines has $\sigma_2\neq0$. 
The phase between the blue and the green lines has $\rho\neq0$.
All other condensates vanish.
}}}}}}
\bigskip\endinsert
\noindent
\midinsert\bigskip{\vbox{{\epsfxsize=3in
        \nobreak
    \centerline{\epsfbox{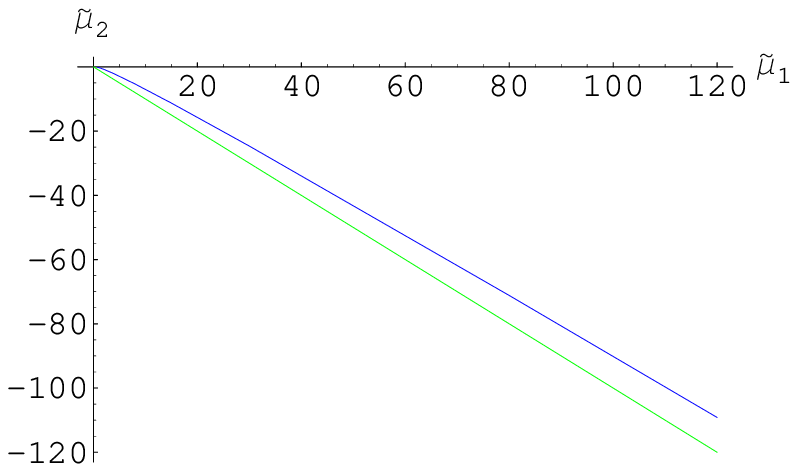}}
        \nobreak\bigskip
    {\raggedright\it \vbox{
{\bf Fig 2.}
{\it  Phase diagram for $LT\ra0$. $\mu_{1,2}$ are horizontal and vertical axes.
Green line $\tmu_2=-\tmu_1$ is a boundary of 
the fundamental domain. 
Blue curve is the line of first order phase transitions
between the phase with $\rho\neq0$ (below) and $\rho=0$ (above). The
leading asymptotics is $\mu_2\approx-\mu_1$.
}}}}}}
\bigskip\endinsert
\noindent
Another possible configuration is the one with partially or
completely restored chiral symmetry.
Consider the phase where $\sigma_1=0,\;\sigma_2\neq0$.
For this configuration we need to use \eomgfb\ with $\p_U X^4=0$
which corresponds to the limit $U_0\ra0$, $\tc\ra\infty$.
Hence, we can use some of the results above in the limit $\tc\ra\infty$.
The value of chemical potential is now
\eqn\cheminfty{  \mu_1= c_1 c^{2\over5} }
which should be compared with \mudiff\ and \fasyla.
The value of $\dS$ is
\eqn\aconeres{  \dS_{1r}=E_0^{7\over2} \left( G^7(0) H(0)+ {2\over7 c_1^{5\over2}} \tmu^{7\over2}\right) }
Analogously, the value of the action for the phase with completely 
restored chiral symmetry is
\eqn\accres{  \dS_{12r}=E_0^{7\over2} {2\over7 c_1^{5\over2}} ( \tmu_1^{7\over2}+\tmu_2^{7\over2}) }
Comparing \aconeres, \accres\ and \sla\ we deduce
the asymptotic behavior of the line of the phase transitions 
between the phase with $\rho\neq0$ and the phase with $\rho=0$.
This line (shown as a blue curve in Figs. 1,2)
must behave asymptotically as $\tmu_2\approx-\tmu_1$ at large values of $\tmu_i$.

More refined features of the phase diagram need numerical work.
The results are summarized in Figs. 1,2. The green lines, $\tmu_2=\pm\tmu_1$
define the boundaries of the fundamental domain.
For $\tmu_{1,2}>\tmu_c$ the phase with straight branes dominates
and all condensates vanish.
Note that $\tmu_c$ is the value of $\tmu_B$ at the
point of the phase transition in the $\tmu_I=0$ plane.
It is computed numerically to be $\tmu_c=0.11$ at $LT\ra0$.
For $-\tmu_c\le\tmu_2\le\tmu_c$ but $\tmu_1$ to the right of the blue line, 
the phase with straight brane ``1'' but curved brane ``2'' (with vanishing electric
field) dominates.
This is the phase with $\sigma_1=0,\;\sigma_2\neq0$.
Below the blue line is the habitat of the curved solution with $\rho\neq0,\;\sigma_1=\sigma_2=0$.
Finally, all condensates vanish between the blue and the red lines, for
sufficiently large values of $\tmu_B$ and $\tmu_I$.

Fig. 2. depicts the phase diagram at a larger scale.
As explained above, the blue phase transition line has
the leading asymptotics $\tmu_2\approx-\tmu_1$.
The correction seems to behave like $\sqrt{\tmu_1}$.

\newsec{Intermediate temperatures}

In this Section we consider the case of finite $LT$.
As we will see, for sufficiently large values of $LT$ the 
phases with nonzero condensates do not exist.
We also assume that the temperature is sufficiently high,
so that gluons are deconfined.
This means that the black hole has formed in the holographic
dual.
The metric is now
\eqn\metrict{    ds^2=\left({U\over R}\right)^{3\over2}(f(U) dt^2+\delta_{ij} dX^i dX^j+(dX^4)^2)
      +\left({U\over R}\right)^{-{3\over2}}[{(dU)^2\over f(U)}+U^2d\Omega_4^2] }
where 
\eqn\defs{ f(U)=1-{U_T^3\over U^3},\qquad U_T={(4\pi)^2 \pi\lambda T^2\over 9} }
The dilaton is given by \dilaton.
The analysis of the previous Section corresponds to the $U_T=0$ limit.
To determine the phase diagram in the $\mu_1-\mu_2$ plane at finite
temperature we need to repeat the analysis of Section 2.

The DBI action now reads
\eqn\dbit{  S=2\int dU U^{5\over2}\sqrt{1-(\p_U A_0)^2+
             f(U) \left({U\over R}\right)^3 (\p_U X^4)^2}  }  
The analogs of \eomtaua\ and \eomgfc\ are
\eqn\eomtauat{ (\p_UX^4)^2={U_0^3 (U_0^5+c^2) R^3 f(U_0)\over U^6 (U^5+c^2) f^2(U)-
   U_0^3 (U_0^5+c^2) U^3 f(U_0) f(U)}
} 
and
\eqn\eomgfct{ (\p_U A_0)^2= {c^2  U^3 f(U)\over U^3 (U^5+U_0^5\tc^2) f(U)-U_0^8(1+\tc^2) f(U_0)}  }
where we again used $c^2=U_0^5 \tc^2$.
It will be convenient to introduce dimensionless variables by rescaling
\eqn\dimless{ U_0= u_0 E_0;\qquad \mu_i=\tmu_i E_0 }
where $E_0=4\pi\lambda/L^2$ as before.

Another useful quantity is 
\eqn\deftt{   \tt={2\pi LT\over 3}    }
Using equations of motion to compute $L$ 
together with \dimless, and going to the rescaled variables,
we obtain the following relation:
\eqn\rell{  \tt=y^{1\over2} G(\tc,y) }
\midinsert\bigskip{\vbox{{\epsfxsize=3in
        \nobreak
    \centerline{\epsfbox{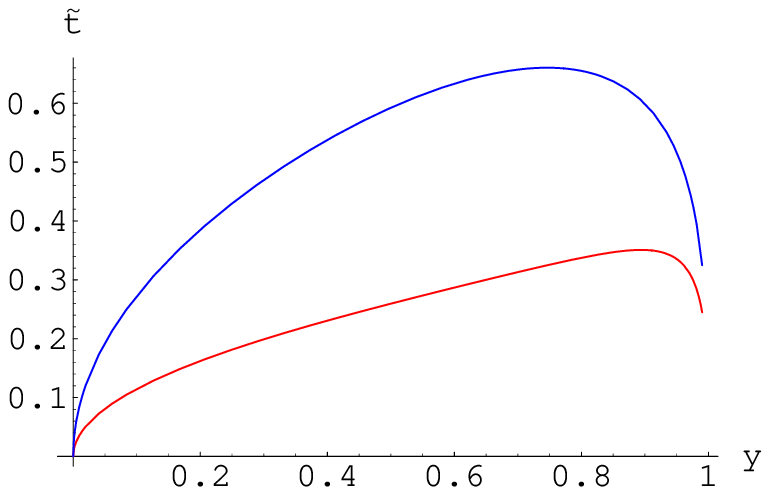}}
        \nobreak\bigskip
    {\raggedright\it \vbox{
{\bf Fig 3.}
{\it  $\tt=y^{1\over2} G(\tc,y)$ [see eq. \rell] as a function of $y$
at $\tc=0$ [red line] and $\tc=1\times10^6$
[blue line]. There is a limiting temperature beyond which the phase with
connected branes does not exists.
}}}}}}
\bigskip\endinsert
\noindent
where 
\eqn\defy{  y=\left({U_T\over U_0}\right)^3={\tt^2\over u_0}  }
where we used \defs, \dimless\ and \deftt, and
\eqn\defbiggt{  G(\tc,y)=\int_1^\infty dx \sqrt{(1+\tc^2) (1-y^2)\over
                (x^5+\tc^2) (x^3-y^3)^2-(1+\tc^2) (x^3-y^3)(1-y^3)}   }
\midinsert\bigskip{\vbox{{\epsfxsize=3in
        \nobreak
    \centerline{\epsfbox{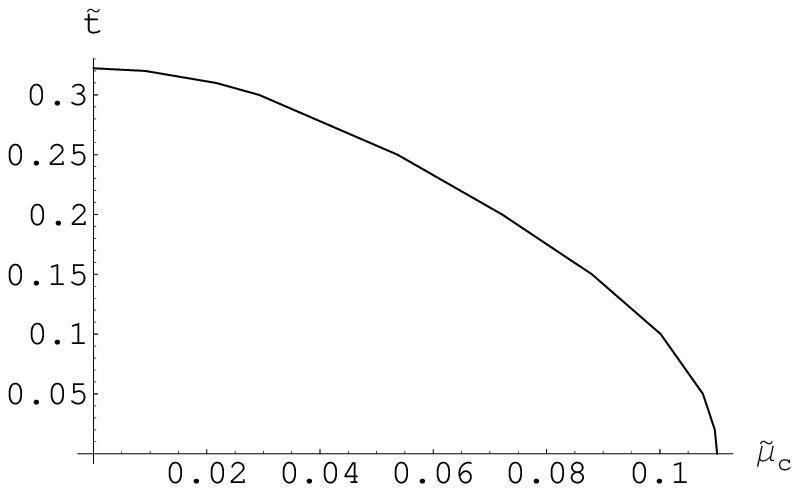}}
        \nobreak\bigskip
    {\raggedright\it \vbox{
{\bf Fig 4.}
{\it  Phase transition line at $\mu_I=0$. Horizontal axis is $\tmu_c$.
Vertical axis is $\tt$.
}}}}}}
\bigskip\endinsert
\noindent
\midinsert\bigskip{\vbox{{\epsfxsize=3in
        \nobreak
    \centerline{\epsfbox{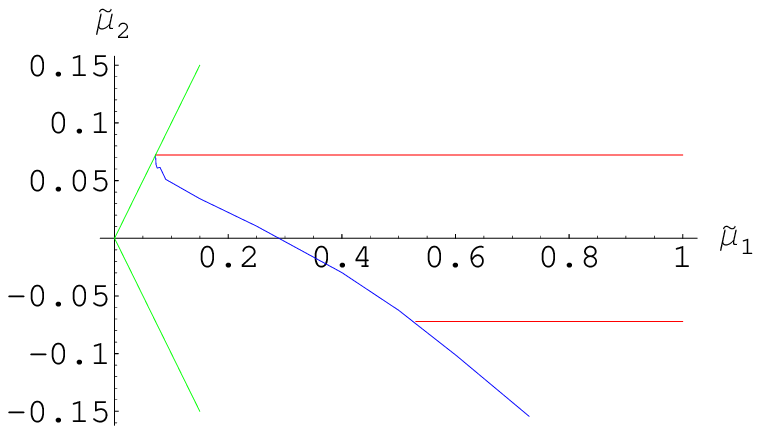}}
        \nobreak\bigskip
    {\raggedright\it \vbox{
{\bf Fig 5.}
{\it  Phase diagram for $\tt=0.2$. 
$\mu_{1,2}$ are horizontal and vertical axes.
Green lines $\tmu_2=\pm\tmu_1$ are boundaries
of the fundamental domain. Solid blue and red curves are the lines of
first order phase transitions.
The phase between the red lines has $\sigma_2\neq0$. 
The phase between the blue and the green lines has $\rho\neq0$.
All other condensates vanish.
}}}}}}
\bigskip\endinsert
\noindent
\midinsert\bigskip{\vbox{{\epsfxsize=3in
        \nobreak
    \centerline{\epsfbox{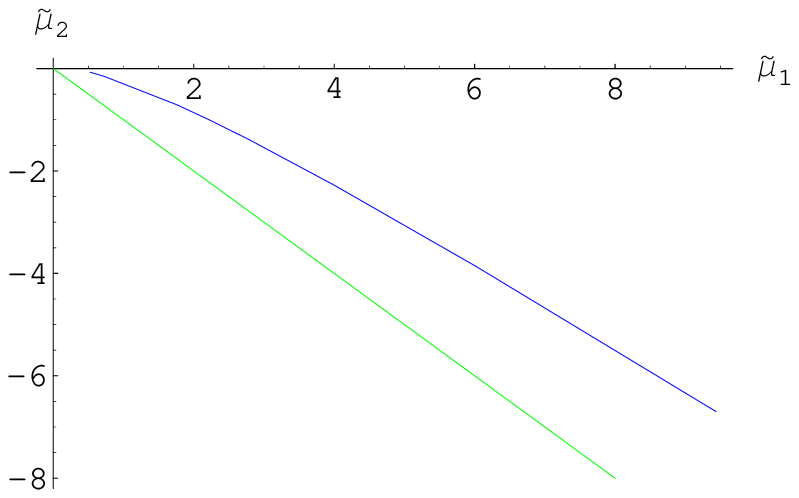}}
        \nobreak\bigskip
    {\raggedright\it \vbox{
{\bf Fig 6.}
{\it  Phase diagram for $\tt=0.2$. $\mu_{1,2}$ are horizontal and vertical axes.
Green line $\tmu_2=-\tmu_1$ is a boundary of 
the fundamental domain. 
Blue curve is the line of first order phase transitions
between the phase with $\rho\neq0$ (below) and $\rho=0$ (above). The
leading asymptotics is $\mu_2\approx-\mu_1$.
}}}}}}
\bigskip\endinsert
\noindent
\midinsert\bigskip{\vbox{{\epsfxsize=3in
        \nobreak
    \centerline{\epsfbox{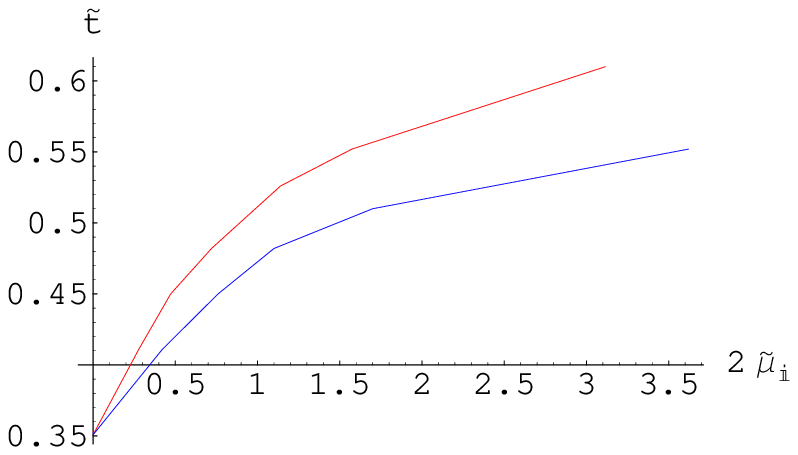}}
        \nobreak\bigskip
    {\raggedright\it \vbox{
{\bf Fig 7.}
{\it  Line of phase transition [shown in blue; bottom line] in the $\mu_B=0$ plane.
Horizontal axis is $2\tmu_I$; vertical axis is $\tt$.
Red [top] line is the limiting value of $\tt$.
}}}}}}
\bigskip\endinsert
\noindent
At a given value of $\tt$ (or, equivalently, $LT$) \rell\
determines $y$ (and $u_0$) as a function of $\tc$.
The behavior of the right hand side of \rell\ is shown in Fig. 3.
There are two solutions of eq. \rell.
The one with larger $y$ comes closer to the black hole horizon
than its smaller $y$ counterpart.
It is the latter one which connects to the vacuum solution at $T=0$
and it is not hard to check that it is thermodynamically preferred.
In the following we will focus on this solution.
Note that for any value of $\mu_I$ there is a limiting temperature
beyond which the curved brane solution does not exists.

The next step is to compute $\tmu_I$
\eqn\relmu{ {1\over2}|\tmu_1-\tmu_2|= {\tt^2\over y}  F(\tc,y)  }
where 
\eqn\defbigfft{ F(\tc,y)=\int_1^\infty dx \sqrt{ \tc^2 (x^3-y^3)\over
          (x^5+\tc^2) (x^3-y^3)-(1+\tc^2)(1-y^3)}     }
Using \rell\ and \relmu\ one can determine $\tc$ and $y$
in terms of $\tmu_{1,2}$.
To determine the thermodynamically preferred state we need 
to compute the action in various phases.
This is again done in a similar way as in the previous Section.
The expression for $\dS$ with non-vanishing $\rho$ is
[compare with \action]
\eqn\actiont{  \dS=2 \left[ \int_{U_T}^{U_0} dU U^{5\over2}+
               \int_{U_0}^\infty dU U^{5\over2}\left(1-
   \sqrt{U^8 f(U)\over U^3 (U^5+c^2) f(U)-U_0^3 (U_0^5+c^2) f(U_0)}\right)\right]  }
In terms of rescaled variables, this becomes
\eqn\actionat{  \dS=2 E_0^{7\over2} {\tt^7\over y^{7\over2}} H(\tc,y)   }
where 
\eqn\htcy{  H(\tc,y)={2\over7} (1-y^{7\over2})+
            \int_1^\infty dx x^{5\over2}\left(1-\sqrt{x^5 (x^3-y^3)\over
                  (x^5+\tc^2) (x^3-y^3)-(1+\tc^2) (1-y^3)}\right) }
Note that the functions $G(\tc,y),F(\tc,y),H(\tc,y)$ defined in
\defbiggt, \defbigfft\ and \htcy\ respectively, reduce
to the functions $G(\tc),F(\tc),H(\tc)$, which were used in
the previous Section, in the limit $y\ra0$.

The value of $\dS$ in the phase with curved branes and
zero electric field ($\sigma_1=\sigma_2\neq0,\;\rho=0$) is
\eqn\snott{  \dS_0=2 E_0^{7\over2} {\tt^7\over y^{7\over2}} H(0,y)   }
where $y$ is determined from \rell\ with $\tc=0$.
To analyze the phase diagram we also need to compute the
value of the action in the phases with $\sigma_i=0,\; \sigma_{j\neq i}\neq0$
and  $\sigma_{1,2}=0$.
The former is given by
\eqn\sonert{  \dS_{ir}={\dS_0\over2}+
    \int_{U_T}^\infty dU U^{5\over2} \left(1-\sqrt{U^5\over U^5+c^2}\right) }
where $c$ is related to $\mu_i$ as
\eqn\ctmui{  \mu_i=\int_{U_T}^\infty dU \sqrt{c^2\over c^2+U^5} }
In terms of the rescaled quantities,
\eqn\sonertres{  \dS_{ir}={\dS_0\over2}+{2\over 7} E_0^{7\over2} \tt^7\left(\sqrt{1+{c^2\over\tt^{10}}}-1
           +\tc^{7\over5} {\Gamma\left({3\over10}\right)\Gamma\left({6\over5}\right)\over\sqrt{5}}
           -{\tc\over\tt^5} {}_2F_1\left({1\over5};{1\over2};{6\over5};-{\tt^{10}\over\tc^2}\right) \right)}
where $\tc$ is determined through
\eqn\tctmutr{  \tmu_i= \tc^{2\over5}\int_{\tt^2\over\tc^{2\over5}}^\infty dx \sqrt{1\over x^5+1}}
Finally, the value of the $\dS$ for the phase with vanishing
condensates is
\eqn\aczeroct{   \dS_{12r}=\dS_{1r}+\dS_{2r}-2 \dS_0}
We are ready to analyze the phase diagram.
As explained at the end of Section 2, $\tmu_c(T)$ defines the line
of phase transitions in the $\tmu_B,T$ plane at $\tmu_I=0$.
Here we reproduce this plot for completeness in Fig. 4.

Figs. 5,6 contain the slice of the 
phase diagram in the $\tmu_1-\tmu_2$ plane for 
a sample value of $\tt=0.2$.
The pictures look similar to the ones in Figs. 1,2.
The only visible difference is shrinking of the $\tmu_c$
with temperature, described by the curve in Fig. 4.
Therefore the picture which emerges at large $\tmu_I$ is the surface of first order
phase transitions in the $\tmu_B,\tmu_I,T$ space whose sections
are the blue lines in Figs. 2, 6.
This surface resembles hyperboloid with $\tmu_I$ being the
axis of rotation.
To determine the behavior of the surface at large $\tmu_I$,
we consider $\tmu_B=0$ plane.

In Fig. 7 we plot the curve which is obtained by intersecting
the phase transition surface with the $\tmu_B=0$ plane. 
Red line on this graph denotes
the limiting value of $\tt$, beyond which the curved brane solution does not 
exist.
Both lines seem to go to limiting values of $\tt$ at large $\mu_I$.
From this picture we see that the phase transition surface 
never intersects the limiting surface, and the transition remains
first order.

\newsec{Conclusions}

In this paper we analyzed the phase diagram of holographic QCD 
at finite temperature and baryon and isospin chemical potentials, defined by \bacp.
The action \mlag\ is invariant under $\mu_1 \leftrightarrow \mu_2$,
which corresponds to $\mu_I\rightarrow -\mu_I$.
Moreover, the physics is also invariant under $\mu_i\ra -\mu_i$.
Hence, the fundamental domain in the $\mu_1,\mu_2$ plane can be taken to be \fdomain.
We found an intricate structure with various phases separated
by the first order phase transitions.\foot{As usual in the DBI analysis,
the free energies of different phases have different values of the derivatives
at the crossing point, indicating first order phase transitions.}.

In particular, at large values of $\mu_I$, we found a surface
of first order phase transitions which separates the phase with the
nonzero value of $\rho$ existing at smaller values of $\mu_B$
from the chirally restored phase at larger values of $\mu_B$.

It is not hard to reinstate gluon confinement/deconfinement
transition in the phase diagram.
It is represented as a surface $T=1/2\pi R_4$.
For temperatures lower than this value chiral symmetry is
necessarily broken and the system is in the phase with 
nonzero electric flux on the branes.
We do not analyze this phase in detail here.
It is clear that in the regions of the phase space where chiral
symmetry is broken, the surface $T=1/2\pi R_4$ describes 
the coincident confinement/deconfinement and chiral phase transitions.

An important observation concerns the energy scales 
in the problem.
We have seen that the relevant variables in the phase 
space are $LT$ and $\mu/E_0$.
This is consistent with the energy scales of the mesons being
$\OO(1/L)$
and constituent masses of quarks being $\OO(E_0)$.

The DBI action used to analyze the structure of the 
phase diagram in this paper is essentially abelian.
It would be interesting to investigate the effect
of non-abelian degrees of freedom, in particular with regards
to the stability of the solutions.
The existence of other phases also cannot be ruled out.
Other directions for future research include generalizing our results
to a higher number of flavors and studying other holographic
models of QCD-like theories.

\newsec{Acknowledgments}
I would like to thank  R. Casalbuoni,  T. Schaefer
and J. Verbaarschot for useful discussions.
I also thank the organizers of the ``Exploring QCD:
Deconfinement, Extreme Environments and Holography''
workshop at Newton Institute, Cambridge, for stimulating
environment.

\footatend\vfill\supereject\immediate\closeout\rfile\writestoppt
\baselineskip=14pt\centerline{{\bf References}}\bigskip{\frenchspacing%
\parindent=20pt\escapechar=` \input refs.tmp\vfill\eject}\nonfrenchspacing
\end